\documentclass[aps,prd,twocolumn,final,nofootinbib]{revtex4}
\usepackage{graphicx} 
\usepackage{amsmath} 
\usepackage[stable]{footmisc}
\usepackage{pseudocode}

\newcommand{\bra}[1]{\left\langle #1 \right|}
\newcommand{\ket}[1]{\left|#1\right\rangle}
\newcommand{\braket}[2]{\left\langle#1 | #2\right\rangle}

\newcommand{\qed}{\nobreak \ifvmode \relax \else
\ifdim\lastskip<1.5em \hskip-\lastskip
\hskip1.5em plus0em minus0.5em \fi \nobreak
\vrule height0.75em width0.5em depth0.25em\fi}

\begin{document}
\title{Quantum Search on the Spatial Grid}
\author{Matthew~D.~Falk}
\affiliation{ MIT 2012, 550 Memorial Drive, Cambridge, MA 02139}
\date{\today} 

\begin{abstract}
\noindent 
This paper explores Quantum Search on the two dimensional spatial grid. Recent exploration into the topic has devised a solution that runs in $O(\sqrt{n \ln{n}})$. This paper explores a new algorithm that gives promise for the $O(\sqrt{n})$ result that is the lower bound off of the grid.
\end{abstract}

\maketitle

\pagestyle{myheadings}
\markboth{M.D.~Falk}{Quantum Search on the Spatial Grid}
\thispagestyle{empty}

\section{Introduction}
\label{Introduction}

Some classical solutions to problems achieve particular speed ups when we allow those algorithms to become quantum based. The most obvious result is the ability to search a group of $n$ elements in sub linear time. Classically, it is required to look at all of the elements, one at a time, to ensure that the marked item is or is not present. In the quantum world, thanks to Lov Grover \cite{FQMADS}, we can do this in $O(\sqrt{n})$ steps and queries. In particular, this subroutine has proved useful for a number of other algorithms and achieving quantum lower bounds. In this paper we explore the complexity of performing this algorithm when reduced to movement on a two dimensional spatial grid. We restrict our model to a quantum robot walking along the two dimensional grid.

We begin by describing the model we adapt for out algorithm in Section \ref{The Model}. We then proceed to a discussion of Grover's Algorithm in Section \ref{Grover's Algorithm} and follow up with the latest results about search on a spatial grid in Section \ref{Quantum Search on the Grid}. Finally, we give our new algorithm for search on the spatial grid with results in Section \ref{Spatial Quantum Search}. We describe the case when there are multiple marked items being search for and how it differs from the non spatial version in Section \ref{Multiple Marked Items} and comment on the implications this gives for other problems in Section \ref{Other Problems}. We give some concluding remarks in Section \ref{Conclusion} and discuss what needs to be done.

\section{The Model}
\label{The Model}

The majority of this paper refers to the "two dimensional grid". This is used to describe a model in which a quantum robot traverses a two dimensional array of points through a quantum walk. That is, a quantum robot is initially placed on the grid at an arbitrary location. They can then move to adjacent nodes, and \textit{only} adjacent nodes in the grid during one time step. At each node, the robot is allowed to read the information that is located there. Our grid consists of all of the elements in $\mathcal X$ that we are interested in looking at. The quantum robot, upon reaching a particular node, can read the value from the grid and perform a query on that element. The robot's movements through the grid are determined by a quantum coin. Whenever the robot moves, it can simultaneously move in all directions (or whichever are described by the coin) in superposition. Thus, we can achieve a superposition of all states in $\mathcal{X}$ in $\sqrt{n}$ steps by performing a walk along the base of the grid and then performing a walk in the perpendicular direction from each of those nodes in parallel.

It is useful to note that each location on the grid can be classified as a two dimensional ket, $\ket{i,j}$, and that the grid is cyclic in both directions. (i.e. $\ket{i,j}$ is connected to $\ket{n, j}$).

\section{Grover's Algorithm}
\label{Grover's Algorithm}

Grover's Algorithm can be viewed as acting on a completey connected graph, $\mathcal{G}_n$, where every node in the graph has an edge connecting it to every other node in the graph. Thus, a quantum robot walking on the graph is able to get to absolutely any node in a single time step. The algorithm is dependent upon this because at every iteration of the diffusion operator each node communicates with every other node in parallel to determine how much amplitude will be transferred between them. 

Herein lies the problem when moving to the grid, nodes can only talk to their direct neighbors. A lot of the edges have been removed from the graph. We can no longer calculate the mean (or invert about it) in a single time step. In order to do this exactly as it is done in Grover's Algorithm, it would require $O(\sqrt{n})$ steps simply to get across the grid and find this information. Thus, we need some other way of performing this calculation. 

The search problem is described as finding a marked item in a given set of values. In this case, we let a ket, $\ket{x}$, represent an element $x \in \mathcal{X}$, our set of all items.
We let the state $\ket{s}$ represent the equal superposition of all elements in $\mathcal{X}$:

\[ \ket{s} = \frac{1}{\sqrt{n}} \sum \limits _x {\ket{x}} \]

Grover's Algorithm is dependent on two subroutines that are repeated $O(\sqrt{n})$ times. The first is the negation unitary. It negates the sole marked item in the set of elements we are looking at. If the marked item is $\ket{w}$, then out negation unitary $U_w$ can be written as:

\[ U_w = I - 2\ket{w}\bra{w} \]

where

\[ U_w^{ \dagger} U_w = \left ( I - 2\ket{w}\bra{w} \right) \left ( I - 2\ket{w}\bra{w} \right) \]

\[ = I - 4\ket{w}\bra{w} + 4\ket{w}\bra{w}\ket{w}\bra{w} = I\]

and 

\[ U_w \ket{w} = \left (I - 2 \ket{w}\bra{w} \right )\ket{w} = \ket{w} - 2 \ket{w} = -\ket{w} \]

\[ U_w \ket{x} = \left (I - 2 \ket{w}\bra{w} \right )\ket{x} = \ket{x} \]

The second subroutine is the diffusion operator: $U_s$.

\[ U_s = 2 \ket{s}\bra{s} - I \]

where unitarity is easily confirmed and 

\[ U_s \ket{s} = ( 2 \ket{s}\bra{s} - I) \ket{s} = 2 \ket{s} - \ket{s} = \ket{s} \]

and

\[ U_s \ket{w} = ( 2 \ket{s}\bra{s} - I)\ket{w} = \frac{2}{\sqrt{n}}\ket{s} - \ket{w} \]

The algorithm consists of repeated applying $U_wU_s$ to the starting state and having the amplitude of the marked state grow with each iteration. 

While this is an optimal algorithm, running as asymptotically fast as it can, the algorithm cannot be translated directly to the spatial grid, due the missing edges of the graph.

\section{Quantum Search on the Grid}
\label{Quantum Search on the Grid}

It was shown by Benioff that Grover's search algorithm took a serious hit when applied to the two dimensional spatial grid. The application still requires $O(n^{1/2})$ queries; however, in between each of those queries, the quantum robot may have to move a distance equal to the diameter of the graph, also $O(n^{1/2})$. Thus, the total running time was $O(n)$ \cite{SSQR}. Aaronson and Ambainis fixed this by giving an algorithm that searches a grid for a single marked item in $O(\sqrt{n} \log ^2 n)$ total steps (queries and walking) \cite{QSSR}. Their algorithm is the main breakthrough in this area and uses Grover's Algorithm recursively on smaller and smaller subcases combined with amplitude amplification.

Later, the search problem on the 2D grid was reduced to a $O(\sqrt{n} \ln n)$ solution by Ambainis et. al \cite{CMQWF}. In 2008, Avatar Tulsi devised a way of improving on the leading algorithm for two-dimensional spatial search using an ancillary qubit. His results lead to an $O(\sqrt{n \ln n})$ solution to find a marked item out of a list of $n$ elements arranged on the vertices of a two-dimensional lattice \cite{FQW}. It remains an open problem whether or not this can be improved upon and the optimal non-lattice solution of $O(\sqrt n)$ reached.

\section{Spatial Quantum Search}
\label{Spatial Quantum Search}

Our algorithm for search on the grid consists of 4 pieces. The first is the building of the superposition on the grid. A quantum robot starting at any location walks along any horizontal in the grid, creates a superposition of all states and then repeats that process along the vertical to create the state $\ket{s}$, stated below for convenience:

\[\ket{s} = \frac{1}{\sqrt{n}} \sum \limits _x {\ket{x}} \]

Once the superposition is created, our robot can act on each node or state simultaneously however information cannot be transferred from one node to another if they are more than a constant number of edges apart. That is, nodes that are $O(\sqrt{n})$ apart cannot communicate and thus we lose the ability to invert about the mean. Our next piece of machinery is to use the negation operator, as Grover did, $U_w$. The next two final pieces of the algorithm are the Local Diffusion operator, $U_{L}^d$, and the Amplitude Dispersion operator, $U_{A}^d$ ($d$ is not a power but a paramter), which are described in the following sections. The algorithm consists of repeated applications of $U_wU_L^4U_wU_A^4$.

\subsection{Local Diffusion Operator}

Even though we cannot talk to all of the nodes on the grid, we are still allowed to learn about our immediate neighbors. We can travel to and communicate with any nodes that live within a constant distance of the current node. However, we cannot talk with a node that is further in communication with a node that we cannot reach. That is, the set of nodes that we are communicating with during a particular time step plus all the nodes that are transitively communicating with us via its communicators must form a constant size set; this to preserve unitarity. We are free to perform any unitary on this subset of nodes. Thus, we can divide the grid into equally sized pieces that tessellate and cover the grid and then perform Grover's Diffusion on each of them. Our diffusion operator must act locally, cover the grid, and preserve unitarity. We constructed our operator in the following form, which allows us to locally spread the amplitudes. It is built up of a superposition of states making small squares that tessellate the entire grid.

\begin{center} $\ket{u_L(i,j)} = \frac{1}{4} \sum \limits _{x,y = 0} ^3 \ket{4i+x,4j+y}$ \end{center}

It is useful to notice that each of the tessellation states are orthogonal to each other and that the states are properly normalized, mainly:

\[ \braket{u_L(i,j)}{u_L(m,n)} = \delta_{i,m}\delta_{j,n}\]

Using this, we can build our unitary operator.

\[ U_L^4 = 2 \sum \limits _{i,j = 0}^{\sqrt{n}/4-1} \ket{u_L(i,j)} \bra{u_L(i,j)} -I \]

Knowing this, it is fairly trivial to show that $U_L$ is unitary.

\begin{tabular}{rcl}
$ {U_L^4} ^{\dagger} U_L^4$ & $=$ & $ (2 \sum \limits _{i,j = 0}^{\sqrt{n}/4-1} \ket{u_L(i,j)} \bra{u_L(i,j)} - I)$ \\
&& $\times (2 \sum \limits _{i,j = 0}^{\sqrt{n}/4-1} \ket{u_L(i,j)} \bra{u_L(i,j)} - I)$ \\
& $=$ & $4\sum \limits _{i,j,m,n = 0}^{\sqrt{n}/4-1} \ket{u_L(i,j)} \braket{u_L(i,j)}{u_L(m,n)}$ \\
&& $ \bra{u_L(m,n)} - 4\sum \limits _{i,j = 0}^{\sqrt{n}/4-1} \ket{u_L(i,j)} \bra{u_L(i,j)} + I$ \\
& $=$ & $4\sum \limits _{i,j,m,n = 0}^{\sqrt{n}/4-1} \ket{u_L(i,j)} \delta_{i,m}\delta_{j,n} \bra{u_L(m,n)}$ \\
&& $ - 4\sum \limits _{i,j = 0}^{\sqrt{n}/4-1} \ket{u_L(i,j)} \bra{u_L(i,j)} + I $ \\
& $=$ & $4\sum \limits _{i,j = 0}^{\sqrt{n}/4-1} \ket{u_L(i,j)} \bra{u_L(i,j)}$ \\
&& $ - 4\sum \limits _{i,j = 0}^{\sqrt{n}/4-1} \ket{u_L(i,j)} \bra{u_L(i,j)} + I$ \\
${U_L^4} ^{\dagger} U_L^4$ & $ =$ & $ I$ \\
\end{tabular}

$U_L^4$ is unitary and therefore we can use it as an operation in our algorithm. The point of this operator is to take local pieces of the grid and trade their amplitudes. That is, if a particular piece of the grid contains the marked item, it will act like Grover's Algorithm on that piece and start sending amplitude to the marked item's node. Otherwise, if there is no marked item in the piece, this operator will actively try to level off the amplitude in that region. 

At this point we introduce the choice of region. We can choose any region that we can tesselate the grid with. For instance, the above unitary operator is based on a region that is a local square. If we let $d$ be the length of the square, we can define a more general, still unitary, operator that acts on the grid:

\begin{center} $\ket{u_L^d(i,j)} = \frac{1}{d} \sum \limits _{x,y = 0} ^{d-1} \ket{di+x,dj+y}$ \end{center}

where we state without proof that

\[ \braket{u_L^d(i,j)}{u_L^d(m,n)} = \delta_{i,m}\delta_{j,n}\]

and

\[ U_L^d = 2 \sum \limits _{i,j = 0}^{\sqrt{n}/d-1} \ket{u_L^d(i,j)} \bra{u_L^d(i,j)} -I \]

with our operator being defined for $d=4$ and $(U_L^d)^{\dagger} U_L^d = I$.

These are just very basic examples of regions that tessellate, the grid. We measure the amount of work a robot has to do for each particular iteration of this operator in terms of $d$. That is, a square region of size $d$ requires the robot to visit $d^2$ nodes on the grid and it would take $O(d)$ steps for the robot to do this and perform the unitary on this region. However, this parameter is chosen ahead of time (even though it is left as a parameter of the system) and is therefore a constant in the analysis of the algorithm. In our version, we choose $d = 4$ for the Local Diffusion Operator, which gives only a constant $O(4) = O(1)$, and negligible, slow down.

The Local Diffusion Operator acts to level out each local region of the grid, unless it contains the marked item. Together with the Negation Operator, $U_w$, this builds the amplitude of the marked item, locally.

\subsection{Amplitude Dispersion Operator}

Simply increasing the amplitude of the marked item within a local region of the grid is not enough. As $n$ may be very large, increasing $\ket{w}$'s amplitude over a finite region will place a very small upper bound on the amplitude that this node can reach. Therefore, we need some way of pulling the amplitude from other regions of the grid. First we notice that by requiring amplitude to move from everywhere on the grid towards our marked item, this puts a lower bound of $O(\sqrt{n})$ on the algorithm, because the robot will need that many steps to reach all other nodes. Our trick here is to perform another diffusion. The second pass over the grid acts very similarly to the previous diffusion operator, except that instead of working on a strictly local region, the new regions are spread out to cover multiple of previous regions. For our algorithm, after trying many regions, we found it extremely successful to make our Amplitude Dispersion Operator the same as the Local Diffusion Operator with a slight shift. The region is a square of size $4 \times 4$ so we decided to move the starting location over 2 nodes in each direction (right and down), in order to get the maximal overlap. Our new states are:

\begin{center} $\ket{u_{A}^d(i,j)} = \frac{1}{d} \sum \limits _{x,y = 0} ^{d-1} \ket{di+x+\lfloor \frac{d}{2} \rfloor,dj+y+\lfloor \frac{d}{2} \rfloor}$ \end{center}

with the operator defined the same way:

\[ U_{A}^d = 2 \sum \limits _{i,j = 0}^{\sqrt{n}/d-1} \ket{u_{A}^d(i,j)} \bra{u_{A}^d(i,j)} -I \]

again with $(U_A^d)^{\dagger} U_A^d = I$, and $d=4$ for our algorithm.

Already unitary by definition of our previous operator, we now have two new tools at our disposal, $U_{L}^d$ and $U_{A}^d$, which provide a slowdown of $d^2$ and act to diffuse and disperse the amplitude respectively.

\subsection{Other Local Regions}

Here we describe some of the other regions that we explored. We measure each each with respect to $d$.

\subsubsection{Four Corners}

This region acts by looking at aset of corners of distance $d$ apart. Similar to the square this takes $O(d)$ steps. These corners provide a smaller amount of computation that needs to be done by the robot, but because it only looks at 4 separate nodes, does not do as well as the $4 \times 4$ square region. However, in our experiments, it performed the same as the $2 \times 2$ square. Our states for 4 corners are: 

\[ u_{c}^d(i,j) = \ket{i,j} \sum \limits _{x,y \in \{0,d\}} \bra{i+x,j+y} \]

\subsubsection{Crosses}

This is the region built up of a node's nearest neighbors in the four cardinal directions. For this region we get $d = 5$, but it only incurs a slowdown proportional to $1$ as it describes the total number of nodes hit, not the diameter of the region and each node is one step away. Our states for the crosses are:

\[ u_{\dagger}(i,j) = \ket{i,j} \left( \sum \limits _{x = \pm 1} \bra{i+x,j} + \sum \limits _{y = \pm 1} \bra{i,j+y} + \bra{i,j} \right) \]

This covers yourself and your nearest neighbors. When tessellation the centers of the next region occur $(2,1)$ away and this ensures that we don't overlap any node more than once on the same pass.

\subsubsection{Restrictions}

One can use any combination of the above regions or define their own. Results will vary based on the tessellation pattern and size of those regions that you choose. We leave it as an open question to find and/or prove the optimal region given a restriction on $d$.

There are many other regions you could choose. However, the larger you go the slower the algorithm performs. It is useful to note that the best performing diffusers are the ones that have the largest number of overlap between the two diffusions. Our algorithm has an average overlap of 4 regions allowing the probability to disperse throughout the grid polynomially fast.
We now introduce a restriction on the choice of $d$, or rather how that choice restricts the rest of the algorithm. In order for the chosen region to properly tessellate the grid, we need both the length and width to be multiples of $d$, or mainly $d^2|n$.

\subsection{Algorithm}

As mentioned previously, our algorithm is similar to that of Grover's; it relies on the repeated application of the unitary.

\begin{enumerate}
\item
Build the superposition over the grid by walking along a horizontal and then a vertical. This takes a one time cost of $O(n^{1/2})$.
\item
Apply the operator $U_wU_{L}^4U_wU_{A}^4$ to the system $O(n^{1/2})$ times. Each iteration takes $O(1 \times 4 \times 1 \times 4)$ steps, as the negation operator is a single step and the others are on the order of $d$.
\item
Measure the state and obtain the marked item with high probability.
\end{enumerate}

\subsection{Results}

As we show next, experiments run with this operator show that the marked item's probability peaks around $n^{1/2}$ iterations; however, we do not get amplitudes arbitrarily close to 1. Most of our amplitudes are on the order of $.70 \rightarrow .95$ which corresponds to a $\frac{1}{2}$ probability or better of measuring the marked item. We are still working on ways to increase this as well as to determine exactly when this maximum amplitude is hit. Either one on its own would prove effective, as once we know how many iterations it takes, we can perform a Local Diffusion Operator on the grid without dispersing. This will build up the marked item as the amplitude that has not reached it yet is stored in the neighboring nodes. This also means that if we measure the node incorrectly, we have been given information as to where the node is and can rerun the algorithm on a smaller portion of the grid in order to find the marked item with even better probability.

Our results are all based on the tessellation of shifted squares of size $4 \times 4$. Below is a graph relating the size of our data $n$ to the number of iterations needed to reach the maximum amplitude and below that the maximum amplitude that is reached.

In every case, the number of required iterations was roughly equal to the square root of the number of items being searched. Thus, we have concluded that the algorithm hits its first peak somewhere within the first $O(n^{1/2})$ iterations; however, we have not proven this rigorously, perhaps it can be done through the use of a geometric display.

The amplitudes decrease as the input size gets larger and larger, which is to be expected as the amplitude is dispersed around a greater number of incorrect nodes. In fact, this algorithm, as the reader can see below, works so as to build up the amplitude around the correct item. That is, it acts as a sink on the grid and pulls amplitudes towards itself. Therefore, the nodes surrounding the marked item will be the most likely options for incorrect measurements.

\begin{figure}[h!]
\centering
\includegraphics[width=8cm,height=7cm]{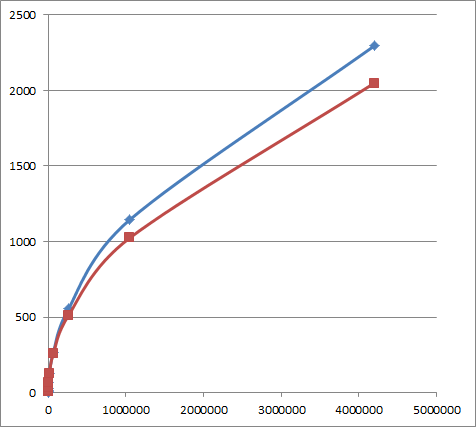}
\caption{This shows the number of iterations required for the marked item to reach its maximum amplitude. Clearly visible in this plot is the trend of the iterations needed to follow $n^{1/2}$.}
\label{Iteration Graph}
\end{figure}

\begin{figure}[h!]
\centering
\includegraphics[width=8cm,height=7cm]{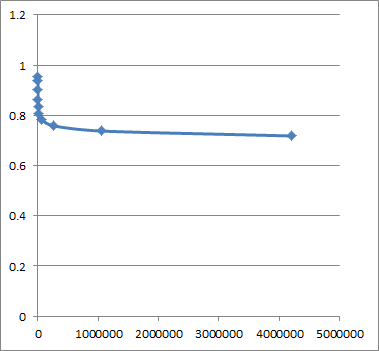}
\caption{This shows the max value of the amplitude of the marked item at the above iteration count.}
\label{Amplitude Graph}
\end{figure}

We suspect that there is a way to perform a final diffusion step on the grid at this final location in order to boost the amplitude of the marked node, but have not found it yet due to the uncertainty in the max amplitude's iteration.

Figures \ref{Iteration Graph} and \ref{Amplitude Graph} show the results of simulating on our algorithm for increasing powers of 4. We present the actual values obtained from out simulations in the following chart:

\begin{center}
\begin{tabular}{cccc}
$n$ & Max Amplitude & Iterations & $\sqrt{n}$\\
\hline
16 & 0.9531 & 2 & 4\\
64 & 0.9373 & 6 & 8 \\
256 & 0.9023 & 12 & 16\\
1024 & 0.8626 & 30 & 32\\
4096 & 0.8338 & 64 & 64 \\
16386 & 0.8073 & 128 & 128 \\
65536 & 0.7812 & 264 & 256 \\
262144 & 0.7581 & 556 & 512\\
1048576 & 0.7377 & 1144 & 1024 \\
4194304 & 0.7178 & 2294 & 2048 \\
\end{tabular}
\end{center}

It is clear that the number of iterations required to hit the maximum aplitude is on the order of $n^{1/2}$. The graph shows that the number of iterations required to get to the maximum aplitude closely follows, while slightly overshooting, $n^{1/2}$. As we can see the maximum amplitude is approaching an asymptote at $\frac{1}{\sqrt{2}}$, which corresponds to a $\frac{1}{2}$ probability of measurement, at the ideal iteration. We suspect that this value can be improved upon, but we will see in the next section the length of time we have in order to try and measure this value.

\subsection{Amplitude Propagation}

While a large maximum probability is nice, it is ideal to have relatively large amplitude for a long period of time (over many consecutive iterations). That way, without knowing exactly how many iterations are needed, one is still likely to measure the actual marked item with high probability. Below we show two different images. They each depict, graphically, the amplitude values for the nodes on a $400$ element set. The first depicts Grover's Algorithm running on this set. The nodes start out in an equal superposition and on each iteration the amplitudes of the wrong elements go down while the marked item's amplitude increases. The cycle is periodic and eventually repeats. The marked element is clearly visible after the very first iteration and stays dominant for most of the progression. Unlike in our algorithm, all of the other nodes decrease equally. Each row of data depicts three separate iterations overlaid next to each other.

The second image depicts our algorithm working on the same set of 400 elements once those elements have been shifted to a $20 \times 20$ spatial grid. Again the nodes start in an equal superposition but now the amplitude travels towards the marked node in waves and also builds up on those nodes that are closest to the marked node. We see a few nodes (unmarked) that have relatively high amplitudes when the marked item peaks, this is not completely bad. If we do in fact measure the wrong item we know that the actual marked item resides fairly close by on the grid and we can repeat the algorithm on a much smaller section of the grid to find the marked item. The marked element is clearly visible after the very first iteration and stays dominant for most of the progression. Unlike with Grover, amplitude travels in waves to the marked node and creates a sink or pyramid around the marked node. First, however, we show some images and describe what is being seen in each of these plots.

\begin{figure}[h!]
\includegraphics[width=6cm,height=4cm]{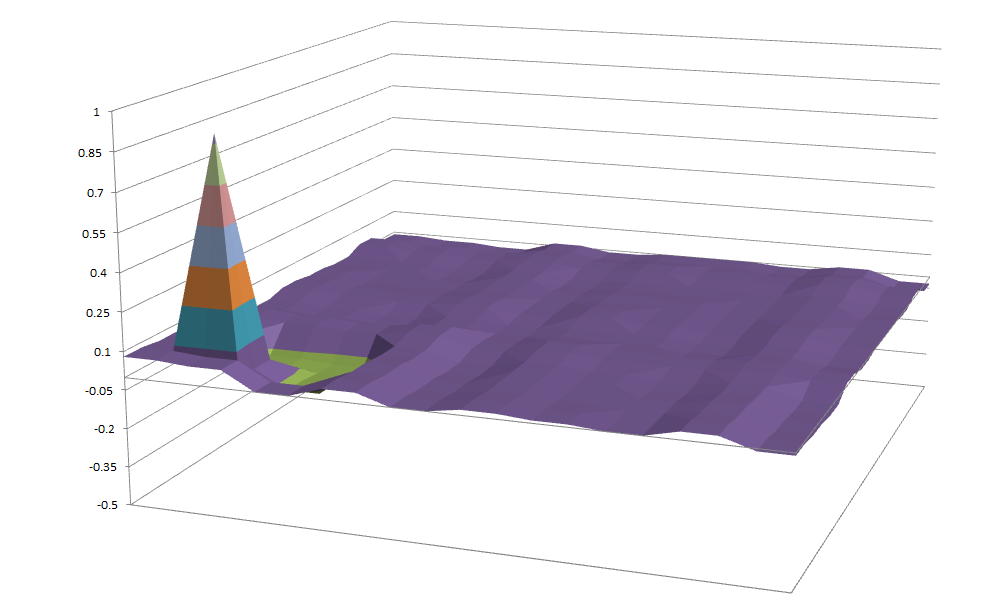}
\caption{This is a close up view of a single iteration of our algorithm.}
\label{Single Iteration}
\end{figure}

In Figure \ref{Single Iteration} we see a close up shot of a single iteration of our algorithm being run on a $20 \times 20$ grid. The different colors represent $.15$ histogram groupings. That is, everything that is a particular color falls within a .15 range of follows. As the graph is of amplitudes, the maximum value is 1. Because none of our values dipped, negatively, below $-0.5$, we decided to cut off the grid there in order to have a closer view of the plot. As we will see on the next page, we have laid out multiple of these images next to each other for convenience, but do not want to confuse what a single iteration looks like.

The iteration depicted is very close to the ideal iteration. We can see that there is a very nice peak indicating the node of the marked item. Additionally, we see a small region where some nodes have a higher amplitude than the rest (the are colored lime green). Here is where most of the amplitude, that is not already part of the marked node, is forming. We have successfully created a sink/pyramid of amplitude around the marked item. Although we have not explored this in too much depth, and thus do not have exact results, this means that when measuring the state, there is an even higher probability (than just that of the marked item) that if we measure a state, it is either correct OR it is a node that is extremely close the desired node. We do not have the numbers corresponding to how close the marked item is, but it is an interesting topic to research. We are very interested in knowing how far away you need to travel from the marked node in order to get a max amplitude that matches Grover's Algorithm off of the grid. While we did not do the study, out intuition is that the distance is directly proportional to $d$, the parameter we use to classify a tesselation's size.

\begin{figure}[h!]
\includegraphics[width=6cm,height=9cm]{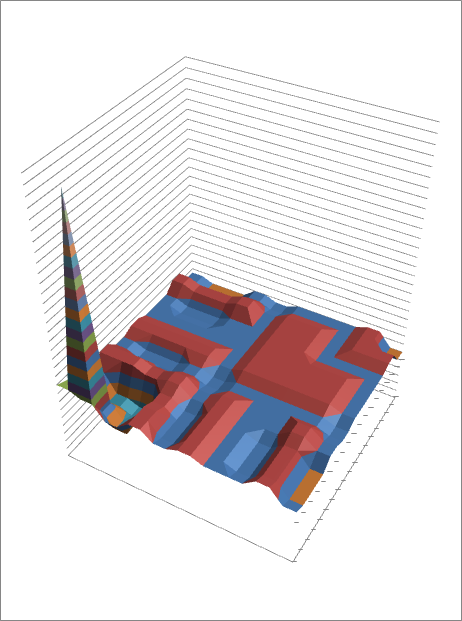}
\caption{This is a close up view of a single iteration of our algorithm, more clearly depicting the amplitude build up around the marked node.}
\label{Amplitude Build Up}
\end{figure}

Below is another image of a different iteration showing more clearly the build up of amplitude that clusters around the marked item. Figure \ref{Amplitude Build Up} depicts an earlier iteration than that of Figure \ref{Single Iteration}, before that amplitude has had enough time to travel all the way to the marked node. 

In the above figure we see the tall spike just as before. However, in this plot in only goes up to $\approx 0.6$ as it is an early stage iteration. What we do see here is that there is a nice sink of light blue nodes surrounding the spike, but only in the direction of its own local tessellation. We also see the wave of amplitude moving horizontally and vertically towards the marked node.

Now we show the actual plots from the simulation of our's and Grover's algorithm side by side.

\newpage

\begin{figure}[h!]
\includegraphics[width=6cm,height=21cm]{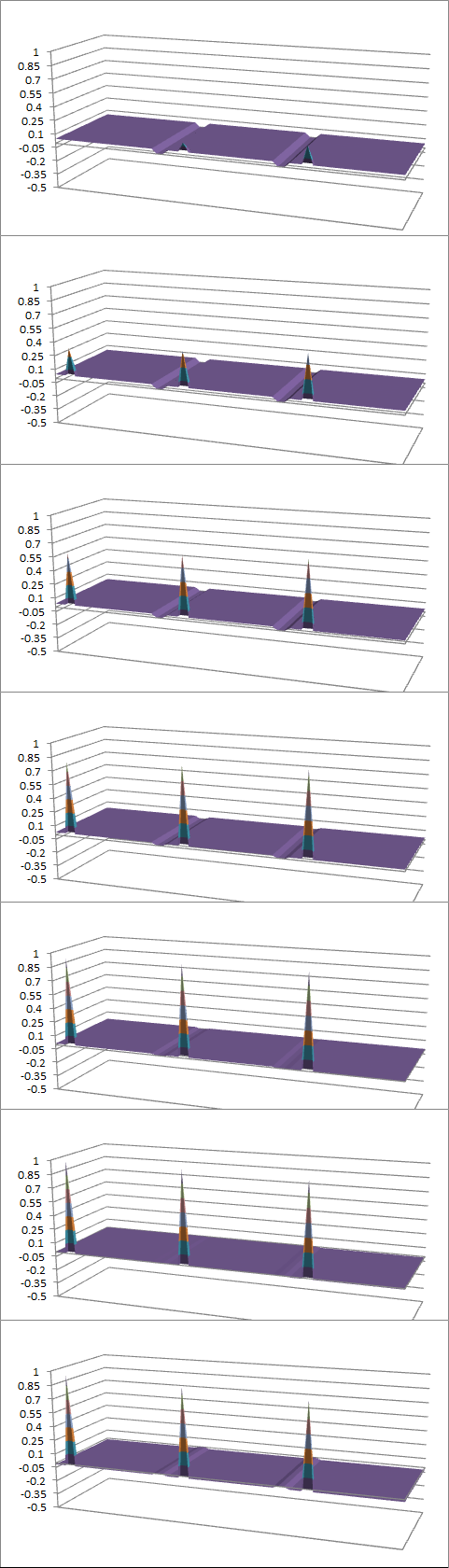}
\caption{This shows the amplitude progression of Grover's Algorithm on a grid of 400 elements.}
\label{Amplitude Progression - Grover's Algorithm}
\end{figure}

\begin{figure}[h!]
\includegraphics[width=6cm,height=21cm]{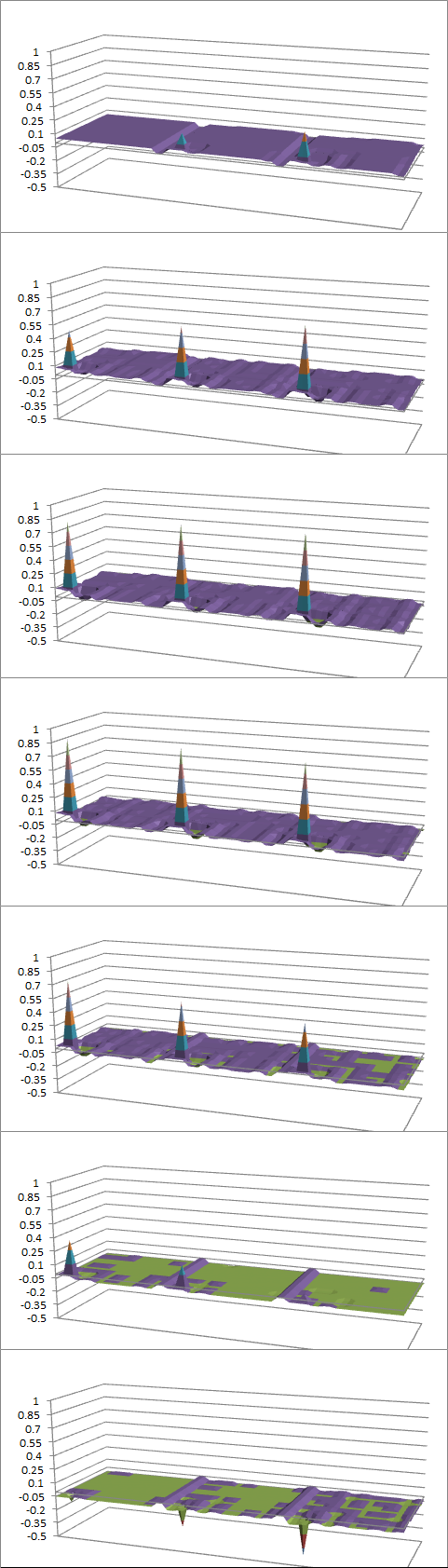}
\caption{This shows the amplitude progression of running our algorithm on a grid of $20 \times 20$ elements.}
\label{Amplitude Progression - Our Algorithm}
\end{figure}

\newpage

\subsection{Comparison}

As seen in Figures \ref{Amplitude Progression - Grover's Algorithm} and \ref{Amplitude Progression - Our Algorithm} above, both algorithms do a great job of isolating the marked item and increasing its amplitude. Our algorithm does it in fewer iterations, which is largely due to our unitary operator which does two rounds on each iteration (one for diffusion and one for dispersal). Each shows the marked item staying prevalent for most of the iterations. While Grover's gets the amplitude to be almost completely singular, our simulation shows that we get the marked item to a high enough amplitude to be measured over a similar time period with high constant probability.

\section{Multiple Marked Items}
\label{Multiple Marked Items}

Here we show the results from running the algorithm when there are multiple marked items. Below we show the results of running the algorithm when there are two marked items. Other than that, the details are the same: same diffusion and dispersion tessellations and still a $20 \times 20$ grid. As expected, the amplitudes don't peak as high, but they shoot up very quickly, much more quickly than when there was a single marked item.

This type of search could be affected differently then the regular search because the marked items' proximity to each other actually matters in the running of out algorithm. Our simulation results show that the closer the marked items are to each other the worse the algorithm does. It still finds and isolates the marked items, but the amplitudes interfere with each other building a larger pyramid of amplitude in the surrounding nodes. When the items are very far apart they each quickly act as if they are the only marked item on their half of the grid. The amplitude differences in these cases were relatively small compared to the rest of the nodes (i.e. 0.62 vs. 0.58). 

We have seen that the total combined probabilities of marked nodes does not stay constant as you increase the number of marked items. In fact, in our above case, it actually decreases. The two nodes had a combined probability of measurement of roughly $72 \%$, while in the single marked item case, at the peak of the algorithm the item had a probability of $79 \%$. We have not explored using different tesselations to boost the multiple marked items avenue, but conjecture that there might be a better tessellation that is independent of the locality of the marked items. This is a great area to explore. We also conjecture that as the number of marked elements (percentage of total elements) increases the algorithm runs in fewer iterations but the combined probability remains roughly equivalent.

\begin{figure}[h!]
\includegraphics[width=6cm,height=12cm]{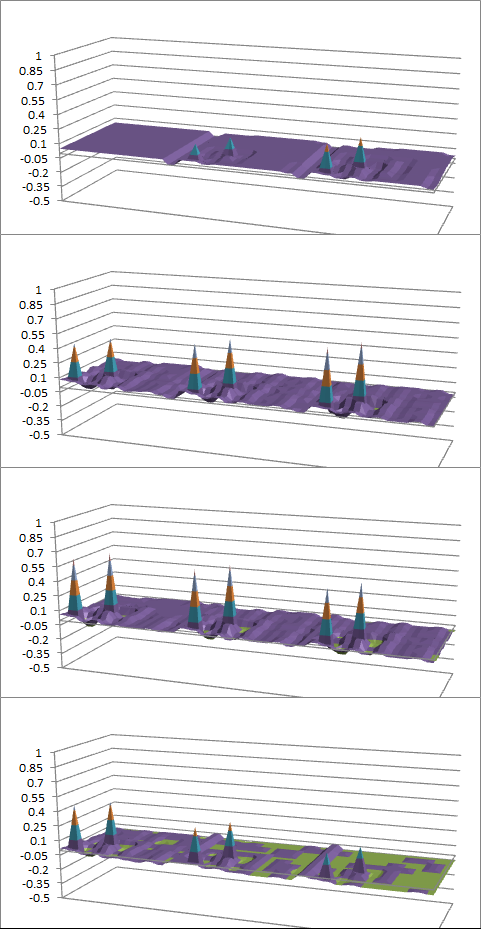}
\caption{This shows the amplitude progression of running our algorithm on a grid of $20 \times 20$ elements when there are two marked elements.}
\label{Amplitude Progression - Multiple Marked items}
\end{figure}

\section{Other Problems}
\label{Other Problems}

If this pans out and proves to be a $O(n^{1/2})$ solution for Search on the Spatil Grid that has many implications for other problems that can be moved to the grid. Mainly, any quantum that relies on searching as its lower bound (with some exceptions) would now be able to be solved on the grid, as well, in the same running time.

For instance, Collision, which has a $O(n^{1/3})$ solution off of the grid would not be affected by moving to the grid. We can still perform the same algorithm: look up $n^{1/3}$ elements and perform a search over a set of $n^{2/3}$ elements. Both on and off the grid would then yield the same complexity, $O(n^{1/3})$.

\section{Conclusion}
\label{Conclusion}

This paper provides the initial outline for what we believe to be the possibility of an optimal Quantum Search algorithm on the Spatial Grid. An interesting question is whether this diffuse and disperse algorithm works in higher dimensions. As we are limited in the above scenario by the diameter of the grid, perhaps in higher dimensions we can improve upon this result. This algorithm would also benefit from Amplitude Amplification, an idea that has not been explored in this paper and determining exactly what the number of iterations needed is. Lastly, we admit that our tessellation pattern is not proved to be optimal, that is, there may be a better dispersion tessellation that couples more nicely with our Local Diffusion Operator.

We have presented the first application of search on a spatial grid that does not rely on Grover's Diffusion Operator in its entirity. We have introduced many new areas for exploration. Mainly: 

\begin{enumerate}
\item
Amplitude amplification from the pyramid of amplitude surround the marked item

\item
Tessellation patterns that increase dispersion

\item
The application into higher dimensions

\item
Tessellation patters that work equally regardless of multiple item locality

\end{enumerate}

These are just some of the new directions that can be explored and should prove some pretty nice results.

\section{Acknowledgments}

I would like to thank Scott Aaronson for the numerous discussion we had about Quantum Algorithms on the Grid and helping me work through new ideas about them. I would also like to thank Vladimir Sobes for helpful discussions about the tessellation patterns and the types of unitary operators we can get on the spatial grid.

\bibliographystyle{plain}
\bibliography{QC}

\begin{thebibliography}{1}

\bibitem{QSSR}
Scott Aaronson and Andris Ambainis.
\newblock Quantum search of spatial regions.
\newblock {\em Proc. 44th Annual IEEE Symp. on Foundation of Computer Science},
  pages 200--209, 2003.

\bibitem{CMQWF}
Andris Ambainis, Julia Kempe, and Alexander Rivosh.
\newblock Coins make quantum walks faster.
\newblock 2004.

\bibitem{SSQR}
Paul Benioff.
\newblock Space searches witha quantum robot.
\newblock {\em Quantum Computation and Information: Contemporary Mathematics},
  305:1--12, 2000.

\bibitem{FQMADS}
Lov Grover.
\newblock A fast quantum mechanical algorithm for database search.
\newblock {\em STOC '96 Proceedings of the twenty-eighth annual ACM symposium
  on Theory of computing}, pages 212--219, 1996.

\bibitem{FQW}
Avatar Tulsi.
\newblock Faster quantum walk algorithm for the two diminsional spatial search.
\newblock {\em The American Physical Society}, 2008.

\end{thebibliography}

\end{document}